# Ion sensitivity of hygroscopic insulator field effect transistors

Joshua N. Arthur*[a,b] and Soniya D. Yambem*[a,b]

[a.] School of Chemistry and Physics, Faculty of Science, Queensland University of Technology (QUT), Brisbane, QLD 4000, Australia
[b.] Centre for Materials Science, Queensland University of Technology (QUT), Brisbane, QLD 4000, Australia
E-mail: joshua.arthur@hdr.qut.edu.au; soniya.yambem@qut.edu.au

Given the many and varied roles of ions in living organisms, biocompatible organic ion sensors are a matter of considerable interest. In this work, simple, low-voltage, solid-state hygroscopic insulator field effect transistors (HIFETs) have been tested to characterise their ion-sensitive properties. Two biologically relevant salt solutions, sodium chloride (NaCl) and potassium chloride (KCl), were tested. To assess pH sensitivity, solutions of hydrochloric acid (HCl) and sodium hydroxide (NaOH) were also tested. The salts and acidic solutions caused similar, concentration-dependent changes in HIFET performance from 10 mM to 1 M, consistent with an increase in ion concentration in the hygroscopic insulator, increasing device capacitance. By contrast, basic solutions caused an overall decrease in device performance, consistent with a net removal of ions due to acid-base reactions between the insulator and the analyte. These results show that HIFETs exhibit promising sensitivity to a range of ions, and can therefore serve as platforms for future ion-selective devices.

## Introduction

Ions play a wide variety of fundamental roles in biological systems. Ion fluxes are extensively involved in vital processes such as cellular signalling, cell volume regulation, and muscle contraction, regulated by selective ion channels and ion pumps in cell walls.[1] Potassium ($K^+$) and sodium ($Na^+$) ions, for example, are involved in nerve impulses and neuron activity. Other biologically relevant ions include protons ($H^+$), calcium ($Ca^{2+}$) and chloride ($Cl^-$).[1,2] Given the vital role of ions in the human body, sensors capable of detecting ion concentration have been researched extensively for health monitoring and biointerfacing applications.[3] Organic thin film transistors (OTFTs), having excellent biocompatibility and signal amplifying properties, are devices particularly suitable for this purpose.[4,5]

OTFT-based ion and pH sensors have been demonstrated using a variety of architectures and sensing mechanisms. This includes organic field effect transistors (OFETs),[6,7] organic electrolyte-gated field effect transistors (EGOFETs),[8,9] and organic electrochemical transistors (OECTs).[10,11] Another class of OTFT with promising characteristics for sensing applications is the hygroscopic insulator field effect transistor (HIFET). HIFETs are simple, solution-processable, all solid-state, low-voltage transistors that utilise the properties of a hygroscopic insulator: poly(4-vinylphenol) (PVP). In its dry, pristine state, PVP is used as an insulator material in OFETs.[12,13] When permitted to absorb moisture, the weakly acidic phenol groups of PVP ionise, resulting in mobile cations ($H^+$). Thus, the hygroscopic layer functions as a solid-state electrolyte, facilitating the low-voltage operation of the device.[14,15] Using the hydrophobic poly(3-hexylthiophene-2,5-diyl) (P3HT) as the active layer, the working mechanism is similar to an EGOFET, where ions accumulate as electrical double layers at the PVP/P3HT interface rather than penetrate and electrochemically dope the active layer.[15] HIFETs can be used as sensors by utilising a permeable top gate electrode, allowing analytes to penetrate and interact with the device. The solid-state architecture permits a self-containment and simplicity in fabrication that is difficult to achieve with electrolyte-based OTFTs, while maintaining the advantages of low-voltage operation. We previously have studied HIFET sensitivity to hydrogen peroxide ($H_2O_2$), and proposed that the primary sensing mechanism is the diffusion of the analyte through the gate and PVP layers, and the oxidation of the active layer.[16-19] There are as yet no detailed studies on the sensitivity of HIFETs to ions.

In this work, we investigate the ion sensitivity of HIFETs using aqueous solutions of sodium chloride (NaCl), potassium chloride (KCl), hydrochloric acid (HCl) and sodium hydroxide (NaOH). We study the response of the HIFET drain-source current ($I_{ds}$) to the deposition of these ionic solutions onto the gate electrode of the device and examine the effect on the figures of merit of the transistor. This study reveals the broad sensitivity of HIFETs to a range of ionic solutions and provides a vital foundation for future work toward optimising and functionalising ion sensitive and ion selective HIFETs.

## Experimental Section

Detailed methods for HIFET fabrication are provided in previous works.[17,19] In brief, glass slides pre-patterned with indium tin oxide (ITO) source and drain electrodes (Xin Yan Technology Ltd.) were used as the substrate. The channel between the electrodes was 50 μm long and 3 mm wide, for a W/L ratio of 60. Approximately 50 nm films of regioregular poly(3-hexylthiophene) (P3HT) (Rieke Metals, LLC, RMI-001EE) were spin coated as the active layer, followed by approximately 700 nm films of PVP (Sigma-Aldrich, 436224) as the hygroscopic insulator. Finally, crosslinked freestanding films of poly(3,4-ethylenedioxythiophene) doped with poly(styrene sulfonate) (PEDOT:PSS) (Heraeus, Clevios PH 1000) were attached to the devices by first depositing 2 μL of 0.1% (3-glycidyloxypropyl)trimethoxysilane (GOPS) solution in deionised water as an adhesion promotor, and lowering the film into place.

The freestanding PEDOT:PSS gates were crosslinked using 1% divinyl sulfone (Sigma-Aldrich, V3700), along with additives 4-dodecylbenzenesulfonic acid (Sigma-Aldrich, 44198) and ethylene glycol (0.1% and 5.8%, respectively).

The PEDOT:PSS solution was deposited into PTFE wells (8 µL per gate) and heated at 70°C for 30 minutes to dry, forming freestanding films.

To test the sensitivity to different ions, 1 M solutions of NaCl, KCl, HCl and NaOH were prepared. HCl was diluted from a 32% stock solution (Ajax Finechem, AJA256). NaOH was prepared in deionized water from pellets (Chem-Supply, SA178). Additional concentrations were prepared by diluting successively with deionized water.

HIFET characterisation was carried out using a Keysight B1500A semiconductor analyzer. Transient $I_{ds}$ sensing responses were recorded by depositing 5 µL of analyte onto the gate electrode above the channel between source and drain electrodes, while applying a constant $V_g$ of -0.3 V and a $V_{ds}$ of -1 V. Forward and reverse transfer sweeps were recorded before and after deposition of the analyte. Gate voltage ($V_g$) was swept at an average rate of ~60 mV s$^{-1}$. Figures of merit were determined from the reverse transfer sweep. Methods for calculating figures of merit are given in previous reports.[19] All figures shown in the report represent averages of typically 5-6 individual HIFETs tested under the same conditions.

# Results and discussion

## HIFET characteristics

In common with previous reports,[14, 20] the HIFETs fabricated for this study utilise a top-gate bottom-contact device architecture, illustrated in cross-section in **Figure 1**A. The active semiconductor layer is P3HT, and the gate electrode is a water-stable and ion-permeable film PEDOT:PSS, crosslinked using divinyl sulfone. Representative output and transfer characteristics, for HIFETs in a pristine condition, are given in Figure 1B and Figure 1C. Performance is comparable to previously reports.[17-19] HIFETs operate within a gate voltage ($V_g$) range of +1 V to -1 V. The large positive threshold voltage ($V_T$) is a result of the polarity of the PVP, which induces charge carriers in the P3HT.[15] By applying a positive $V_g$, cations from the PVP are caused to accumulate at the P3HT/PVP interface, counteracting the polarity effect and de-doping the channel to reach an "off" state.

## Transient $I_{ds}$ modulations

In an OTFT sensor, the analyte must interact with the device in such a way as to cause a detectable change in its electrical characteristics. The most direct means to probe the characteristics of an OTFT is to monitor the change in $I_{ds}$ over time, under fixed voltages. In **Figure 2**, we show the transient change in $I_{ds}$ in HIFETs upon the deposition of 5 µL of different ionic analytes onto the gate electrode. We plot the percent change in $I_{ds}$ relative to the $I_{ds}$ measured immediately before deposition of the analyte, at t=0 seconds (100×$\Delta I_{ds}/I_{ds}$(t=0)). During the experiment, we fix the drain-source voltage ($V_{ds}$) and $V_g$ at -1 V and -0.3 V, respectively. This corresponds to the saturation regime, with a moderately doped (already 'on') channel, which we find maximises the $I_{ds}$ modulation (see **Figure 5**). To quantitatively evaluate these results, we extract the maximum $I_{ds}$ modulation reached within the first 200 seconds of exposure to the analyte, as well as a sensing parameter β, defined in Equation 1. The β parameter represents the quantity of charge that passes through the channel ($Q = \int I \, dt$) that is gained due to the analyte. This is given as a fraction of the charge that would have passed through the channel had $I_{ds}$(t=0) remained constant.

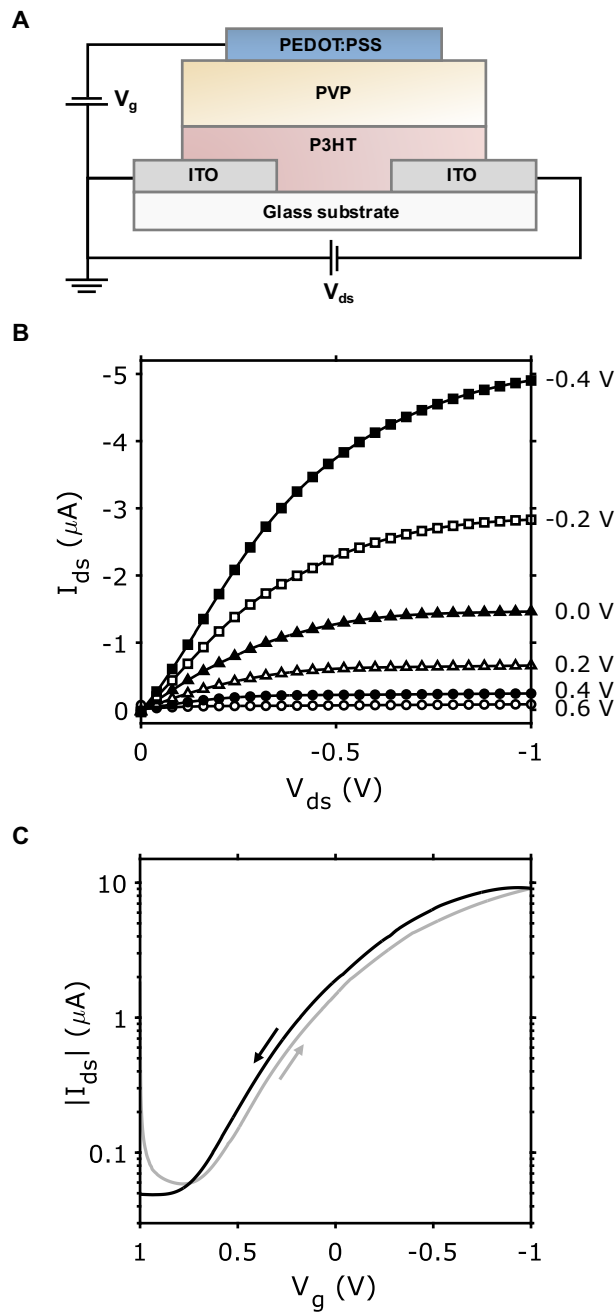

**Fig. 1 (A)** HIFET cross-sectional layer structure. **(B)** Representative HIFET output characteristics, measured at various gate voltages (indicated beside each curve). **(C)** Representative HIFET transfer characteristics, showing forward and reverse sweeps ($V_{ds}$ = -1 V).

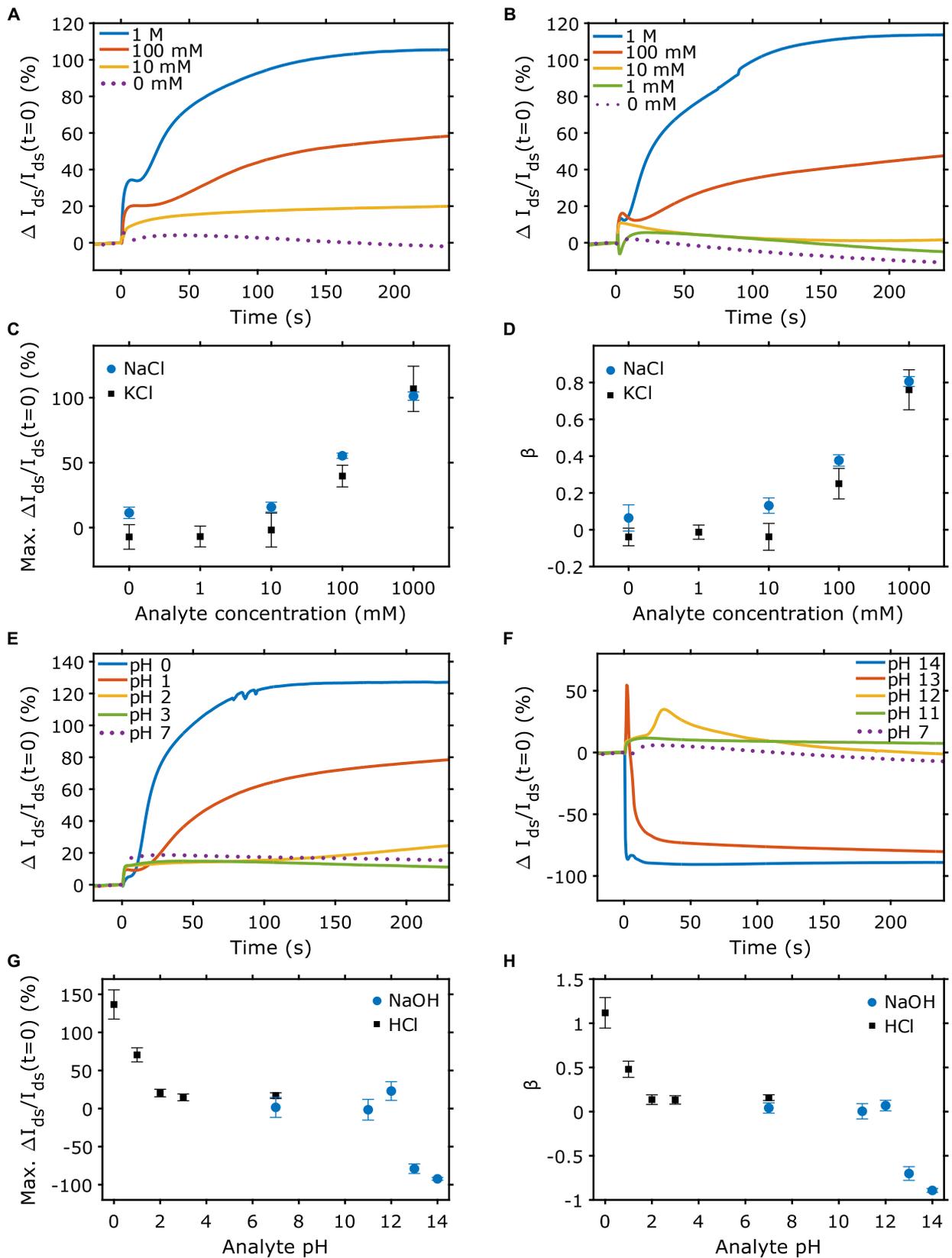

**Fig. 2** The transient change in $I_{ds}$ upon deposition of **(A)** NaCl solutions, and **(B)** KCl solutions, of different concentrations. Trials from representative devices are shown (see Figure S1 and Figure S3 for full data). The response to deionised water (0 mM) is shown as a dotted line. For both NaCl and KCl, the average **(C)** maximum $I_{ds}$ modulation, and **(D)** sensing parameter β, are shown. Error bars indicate the standard deviation. **(E)** and **(F)** show the transient change in $I_{ds}$ upon deposition of HCl (pH 0-3) and NaOH (pH 11-14) solutions. Deionised water is listed as pH 7. Trials from representative devices are shown (see Figure S4 and Figure S5 for full data). **(G)** and **(H)** plot the maximum $I_{ds}$ modulation, and sensing parameter β, for the full range of pH values tested.

$$\beta = \frac{\int_{t=0}^{t=200}[I_{ds}(t)-I_{ds}(0)]dt}{\int_{t=0}^{t=200}I_{ds}(0)dt} \qquad 1$$

Figure 2A and 2B show representative $I_{ds}$ responses to NaCl and KCl, respectively. These are chosen from multiple trials on different devices, which are provided in full in Figure S1 and S3 in the supporting information. Average maximum modulations and β parameters are given in Figure 2C and 2D. The response to deionised water (0 mM) is included to show the "background" signal due to the effect of hydration on the device. HIFETs exhibit a strong concentration-dependant response to both NaCl and KCl solutions. For 1 M solutions, $I_{ds}$ increases by over 100%. A positive modulation reflects an increase in the magnitude of the hole current between source and drain, and hence increased charge in the P3HT channel. The transient $I_{ds}$ is slow to reach a maximum, and proceeds in two stages, the first being smaller and faster. The response time is related to the initial hydration state of the device. Where the device is pre-hydrated with deionised water prior to testing, the $I_{ds}$ modulation is more rapid, while reaching a similar maximum (see Figure S2).

Overall, the sensitivity of HIFETs to both salts is similar, indicating that HIFETs are sensitive to both $Na^+$ and $K^+$. Average modulations for KCl are generally slightly smaller, but the smaller response is also evident in the background signal to due to hydration. This suggests there was a difference in the initial state of hydration in both batches of devices (fabricated and tested separately), which can be expected to vary with ambient room humidity. At 10 mM and below, modulations begin to become indistinguishable from the background response due to hydration.

Modulation in $I_{ds}$ for HCl and NaOH solutions is plotted in Figure 2E to 2F. Concentrations of HCl and NaOH are given in terms of calculated pH, where pH 0 corresponds to 1 M HCl, and pH 14 to 1 M NaOH. Deionised water, indicating the background hydration effect, is plotted as pH 7. There is a positive $I_{ds}$ modulation in response to highly acidic solutions (Figure 2E) and a negative modulation in response to strongly basic solutions (Figure 2F). The sensitivity to HCl corresponds closely to NaCl and KCl. Average modulations are generally larger and faster, likely again related to the initial hydration state of the transistors, but also possibly reflecting the increased mobility of the smaller cation ($H^+$).[21] However, the sensitivity limit is similar, where HCl solutions of pH 2 (10 mM) are difficult to distinguish from the background response to hydration. The negative modulations observed for strongly basic NaOH solutions (pH 13 and 14) indicate a different mechanism from the other analytes tested; the analyte having the effect of decreasing charge in the channel. Like the other analytes, the modulation becomes difficult to distinguish from the effect of water at 10 mM (pH 12).

**Effect of salts on HIFET figures of merit**

To gain deeper insight into the effect of ions on HIFET performance and the underlying mechanisms, changes in key figures of merit after deposition of the analyte were examined. **Figure 3** plots the average ON/OFF ratio, threshold voltage ($V_T$), transconductance ($g_m$), and the product of mobility and capacitance ($\mu_{sat} \times C$), extracted from the transfer characteristics of HIFETs before and after depositing KCl solutions of different concentrations. Additional figures of merit for KCl and the same for NaCl are given in the supporting information (see Figure S6 and Figure S7).

The ON/OFF ratio is the ratio of the maximum to minimum $I_{ds}$ recorded in the transfer curve. The maximum $I_{ds}$, or the 'ON' current occurs at $V_g$ = -1 V, while the minimum or 'OFF' current occurs around $V_g$ = +0.8 V. In Figure 3A, the ON/OFF ratio increases with increasing KCl concentration. At 1 M, the change is dramatic: the average ON/OFF ratio increasing from 90±10 to 310±50. This is largely the result of increased ON currents (Figure S6A), while OFF currents were not significantly changed (Figure S6B). Similar behaviour is observed for NaCl, showing significant increase in ON currents with analyte concentration (Figure S7A), but more variation in the OFF currents (Figure S7B) led to less concentration-dependence in the ON/OFF ratio (Figure S7C). Overall, this behaviour reflects an improvement in the ability of the transistor to accumulate charge in its ON state as ion concentration increases. This effect is reflected clearly in the other figures of merit.

Threshold voltage is strongly dependant on ion concentration (Figure 3B). We determine threshold voltage by the conventional method derived from the gradual channel approximation.[22] Ideally, $V_T$ represents the $V_g$ at which mobile charge begins to accumulate in the channel, or from the opposite perspective, the $V_g$ at which the mobile charge is first depleted. As ion concentration increases, $V_T$ shifts significantly from around +0.5 V toward 0 V. This shows that additional ions inside the HIFET help the device to more efficiently (at smaller $V_g$) deplete the channel.

From Figure 3C, the transconductance ($g_m$) is also strongly dependent on ion concentration, which is directly related to the shift in $V_T$. In Figure 3C, we plot the maximum $g_m/W$, defined as the maximum of the derivative of the transfer sweep ($g_m = \partial I_{ds}/\partial V_g$) normalised by channel width (W = 3 mm). This characterises how well a change in $V_g$ can control charge density in the channel and modulate $I_{ds}$. Hence, for a greater transconductance, a larger $I_{ds}$ can be achieved in the 'ON' state, increasing the ON/OFF ratio, and a smaller positive $V_g$ is needed to deplete mobile charge from the channel, shifting $V_T$.

Figure 3D shows the product of saturation mobility ($\mu_{sat}$) and capacitance (C), calculated using the Equation 2, which is derived from the gradual channel approximation.[23] Here, *L* and $W$ are the channel length and width, respectively, and the value $\partial\sqrt{I_{ds}}/\partial V_g$ is approximated as the gradient of a line fitted to the plot of $\sqrt{I_{ds}}$ vs $V_g$. This parameter, $\mu_{sat} \times C$, increases with ion concentration, and reflects the same phenomenon as $g_m$. The increase in $\mu_{sat} \times C$ is most likely due to an increase in total device capacitance with ion concentration, as the hole mobility, $\mu_{sat}$, of the P3HT is unlikely to be affected by ion concentration. An increase in capacitance with ion concentration is consistent with the Gouy-Chapman-Stern model, which describes the double layer capacitance at an electrode/electrolyte boundary as proportional to the square root of the ionic strength of the electrolyte (ionic strength is equivalent to concentration for KCl and NaCl).[24] This relationship roughly agrees with the trend shown in Figure 3C. Experimentally, the double layer capacitance of a P3HT-based EGOFET has been shown to vary in this way with the KCl concentration of the electrolyte.[24] Thus, it is likely that the primary mechanism for $I_{ds}$ modulation in response to KCl and NaCl is an increase in the total capacitance of the HIFET. Ions from the solution deposited onto the gate electrode diffuse through the permeable PEDOT:PSS gate and enter the PVP layer, adding to the existing ions and increasing the total ionic strength. This is fundamentally different to the $H_2O_2$ sensing mechanism, explored in our previous works.[18, 19] There, $H_2O_2$ directly oxidises the P3HT channel, decreasing ON/OFF ratio (due to increased OFF currents), increasing rather than reducing $V_T$, and having no effect on $\mu_{sat} \times C$.[19]

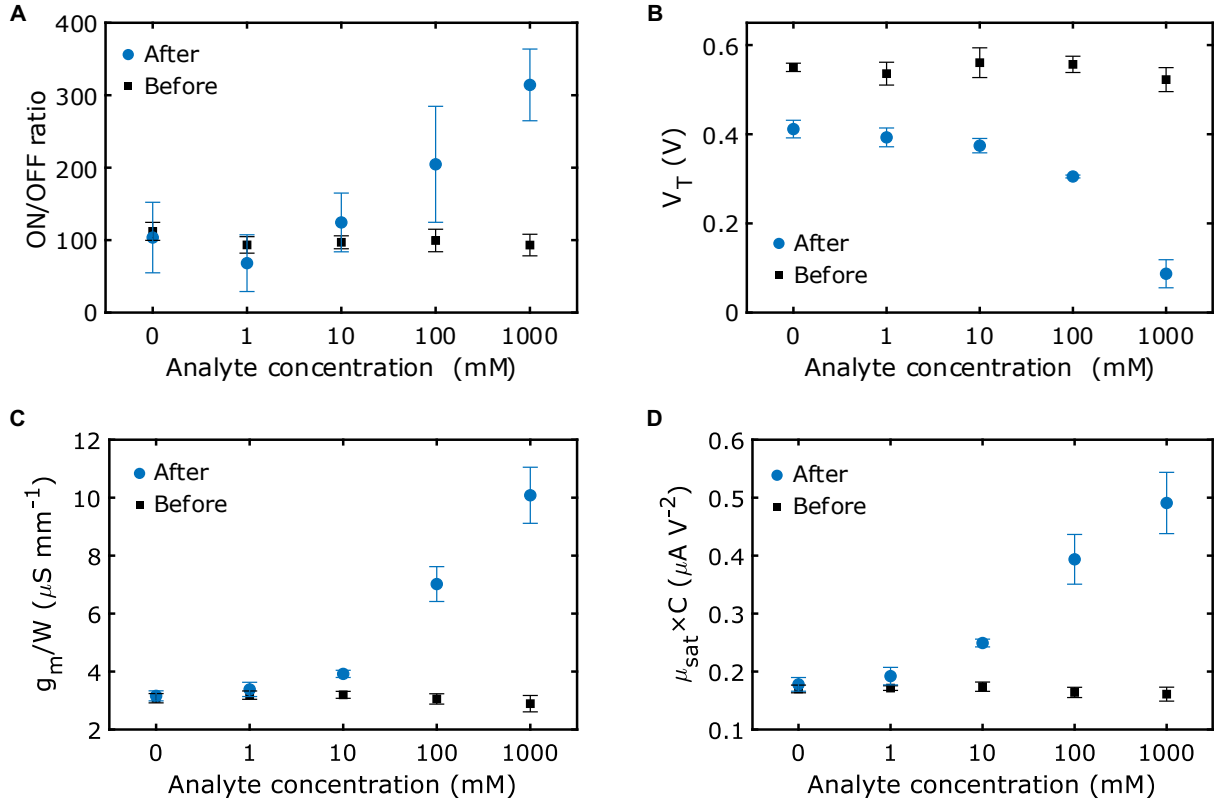

**Fig. 3** Average figures of merit for HIFETs in the saturation regime before and after deposition with KCl solutions of different concentrations. **(A)** ON/OFF ratio. **(B)** Threshold voltage ($V_T$). **(C)** Maximum transconductance ($g_m$). **(D)** Product of saturation mobility ($\mu_{sat}$) and capacitance (C). Other major figures of merit for KCl are given in Figure S5. Figures of merit for NaCl are given in Figure S6.

$$\mu_{sat} \times C = \frac{2L}{W}\left(\frac{\partial\sqrt{I_{ds}}}{\partial V_g}\right) \qquad 2$$

It is noteworthy that the figures of merit, particularly $\mu_{sat} \times C$, of HIFET sensors are highly sensitive to ion concentration, with good consistency between trials (as indicated by the small standard deviations). Unlike the modulations and β parameters extracted from the transient $I_{ds}$ response (Figure 2E and Figure 2F), $\mu_{sat} \times C$ can clearly distinguish between the effect of deionised water and 10 mM KCl or NaCl. Even at 1 mM, $\mu_{sat} \times C$ is clearly different to water, though the standard deviations overlap. This places HIFET ion sensors in a biologically relevant sensitivity range. In human blood serum, for example, a range of 3.5-5.5 mM for $K^+$ and 135-145 mM for $Na^+$ is considered normal, and serious disease is indicated by levels outside these ranges.[25, 26]

**Effect of pH on HIFET figures of merit**

Key figures of merit for HIFETs before and after deposition of HCl and NaOH solutions of varied pH are shown in **Figure 4**. HCl solutions from 1 M to 1 mM are plotted as pH 0 to 3, and NaOH solutions from 1 mM to 1 M are plotted as pH 11 to 14. As the acidic and basic solutions were tested in separate batches, there is a separate statistic for pH 7 (deionised water) corresponding to each.

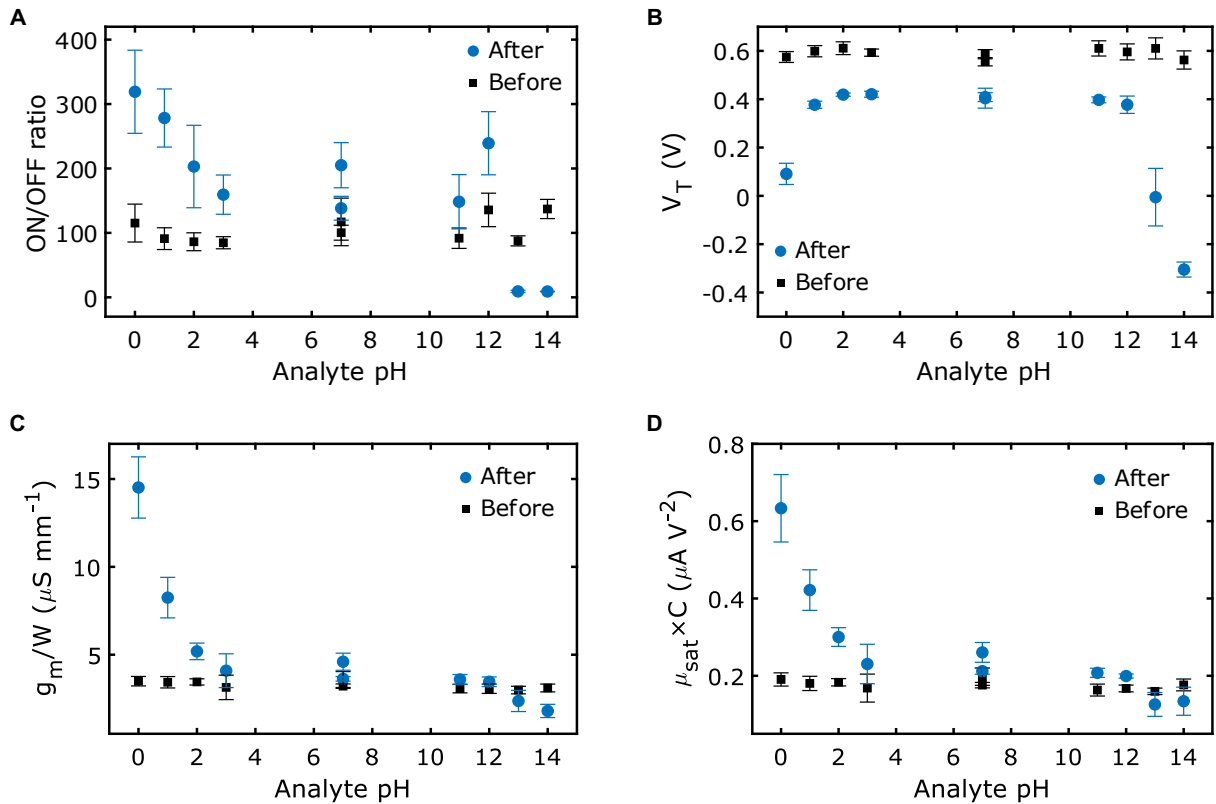

**Fig. 4** Average figures of merit for HIFETs in the saturation regime before and after deposition with HCl and NaOH solutions of different pH. **(A)** ON/OFF ratio. **(B)** Threshold voltage ($V_T$). **(C)** Maximum transconductance ($g_m$). **(D)** Product of saturation mobility ($\mu_{sat}$) and capacitance (C). Other major figures of merit for HCl and NaOH are given in Figure S7.

The effect of acidic, HCl solutions, is comparable to the effect of NaCl and KCl. The ON/OFF ratio increases with HCl concentration (Figure 4A), $V_T$ shifts towards 0 V (Figure 4B), and both $g_m$ (Figure 4C) and $\mu_{sat} \times C$ (Figure 4D) increase. At pH 0 (1 M), HCl consistently causes slightly larger changes than 1M NaCl or KCl, but there are otherwise no indicators of distinct behaviour owing to the $H^+$ ion. It is likely that the capacitance mechanism proposed above also accounts for the sensitivity to low pH. We also note that, like the salt solutions, $\mu_{sat} \times C$ is more sensitive to pH than statistics derived from the transient $I_{ds}$ modulation. HCl solutions of pH 2 are distinguishable from pH 3 and pH 7.

The effect of basic, NaOH solutions, clearly exhibits a different underlying mechanism. NaOH solutions of pH 13 and 14 significantly decrease the ON/OFF ratio. An examination of the ON and OFF currents (see Figure S8A and S8B) shows that this is a result of both a decrease in ON current and increase in OFF current. $V_T$ shifts in the negative direction with increasing pH, and at the same time there is a decrease in both $g_m$ and $\mu_{sat} \times C$. These results are consistent with an overall decrease in capacitance in response to highly basic solutions. Before deposition of an analyte, the capacitance of the pristine HIFET is due to the concentration of mobile $H^+$ and corresponding anions within the layer, owing to the presence of moisture and the weakly acidic properties of the phenol groups along the PVP polymer chains. We propose that deposition of NaOH solutions results in an influx of $OH^-$ ions that would rapidly neutralise available $H^+$, producing $H_2O$, and $Na^+$ would react with the ionised phenol groups (phenoxide) to form sodium phenoxide.[27] The result would be fewer ions in the PVP layer to participate in the formation of electrical double layers, and hence lower capacitance and device performance.

Our results demonstrate that HIFETs, as fabricated here, are only sensitive to the extreme ends of the pH scale. At low analyte concentrations, from pH 3-12, the effect on HIFET figures of merit is not distinguishable from that of deionised water. This means that the sensitivity of HIFETs to other ions will not be significantly affected by normal background pH levels in the solution (for example, blood has pH level range of 7.35 to 7.45).[28] Additionally, HIFETs will have an application in sensing extreme levels of pH (for example, human gastric acid is between pH 1 and 2.5).[29]

**Effect of $V_g$ on the $I_{ds}$ modulation**

**Figure 5** shows the effect of $V_g$ on the transient $I_{ds}$ modulations upon deposition of 1 M NaCl, while $V_{ds} = -1$ V. In our earlier study of HIFET $H_2O_2$ sensitivity, we noted an apparent relationship between the size of the $I_{ds}$ modulation and the initial charge density of the channel. That is, where device voltages caused an initially low charge density, the $I_{ds}$ modulation was larger.[19] This is not the case for ionic analytes. Instead, maximum modulations grow with increasingly negative $V_g$ up to -0.3 V. This is consistent with the mechanism we have described. A more negative $V_g$ ensures that an increase in capacitance will increase the doping in the channel. At -0.6 V, we observe that $I_{ds}$ decays rapidly over time. This behaviour was observed in the $H_2O_2$ sensitivity study.[19] This consistent pattern of instability in the transient $I_{ds}$ at larger, negative $V_g$, may point to underlying non-ideal properties of the transistor. We note that the gradient of the transfer curve (Figure 1C) is significantly reduced in this region, indicating the same limitation.

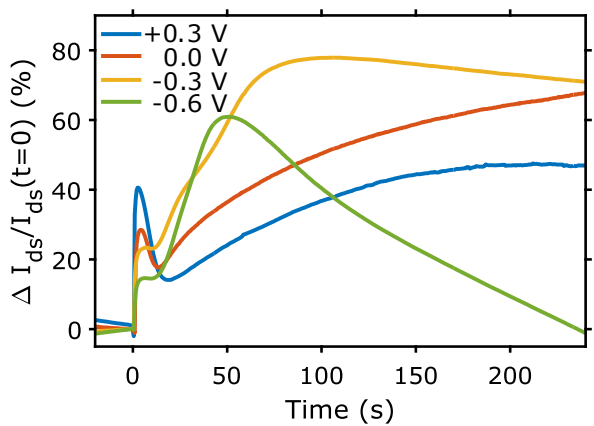

**Fig. 5** The transient change in $I_{ds}$ upon deposition of 1 M NaCl solutions, while applying different $V_g$ (+0.3 V, 0.0 V, -0.3 V, -0.6 V). $V_{ds}$ is fixed at -1 V in all trials. Trials from representative devices are shown (see Figure S8 for full data).

## Conclusion

We have demonstrated HIFETs to be sensitive to a range of ionic solutions, NaCl, KCl, HCl, and NaOH, indicating a broad range of possible applications. Modulations in $I_{ds}$ and key figures of merit are highly dependent on concentration as well as the nature of the ionic solution (acidic or basic) and we have discussed the mechanisms of these changes. We proposed that the positive modulation in $I_{ds}$ for ionic solutions is due to the diffusion of the ions into the hygroscopic insulator, PVP, increasing the total capacitance of the device. The negative effect of NaOH on $I_{ds}$ is consistent with acid-base reactions between NaOH and the phenol groups of PVP, reducing the net ion concentration and device capacitance. Among the key figures of merit $\mu_{sat} \times C$, is more sensitive to concentration and can detect as low as 1mM concentration for KCl.

In summary, HIFETs are platform devices with great potential to become effective ion sensors, by both improving sensitivity and achieving specific ion selectivity. In the future, the HIFETs can be made ion selective by incorporating ion-selective membranes[30, 31] above the gate electrode. There are also number of available means to achieve improved sensitivity. As noted, pre-hydrating the device can increase the response time of the $I_{ds}$ modulation. In many biological applications, the device will be operated in a consistently hydrated environment, so achieving a consistent background state is unlikely to be a significant issue. Maintaining the device in a fully hydrated state prior to exposure to the analyte will also eliminate the background response to water.

On the basis of the proposed ion sensing mechanism, we also suggest that decreasing the initial capacitance of the device may help to improve the sensitivity at low ion concentrations. For example, pre-treating the PVP film to reduce the initial mobile $H^+$ concentration. Further, overall device capacitance could also be modified by altering the gate electrode. We have previously observed that $\mu_{sat} \times C$ is highly dependent on conductance of the electrode.[32] Therefore, HIFETs offer great potential for ion sensing applications and our work will be foundational for future developments in this area.

## Author Contributions

## Conflicts of interest

## Acknowledgements

J. N. A. is supported through Australian Government Research Training Program Scholarships. Device fabrication and testing was performed at the Central Analytical Research Facility (CARF) and supported by Faculty of Science.

# Ion sensitivity of hygroscopic insulator field effect transistors

*Joshua N. Arthur,[a,b] Soniya D. Yambem[a,b]*

[a]School of Chemistry and Physics, Faculty of Science, Queensland University of Technology (QUT), Brisbane, QLD 4000, Australia
[b]Centre for Materials Science, Queensland University of Technology (QUT), Brisbane, QLD 4000, Australia

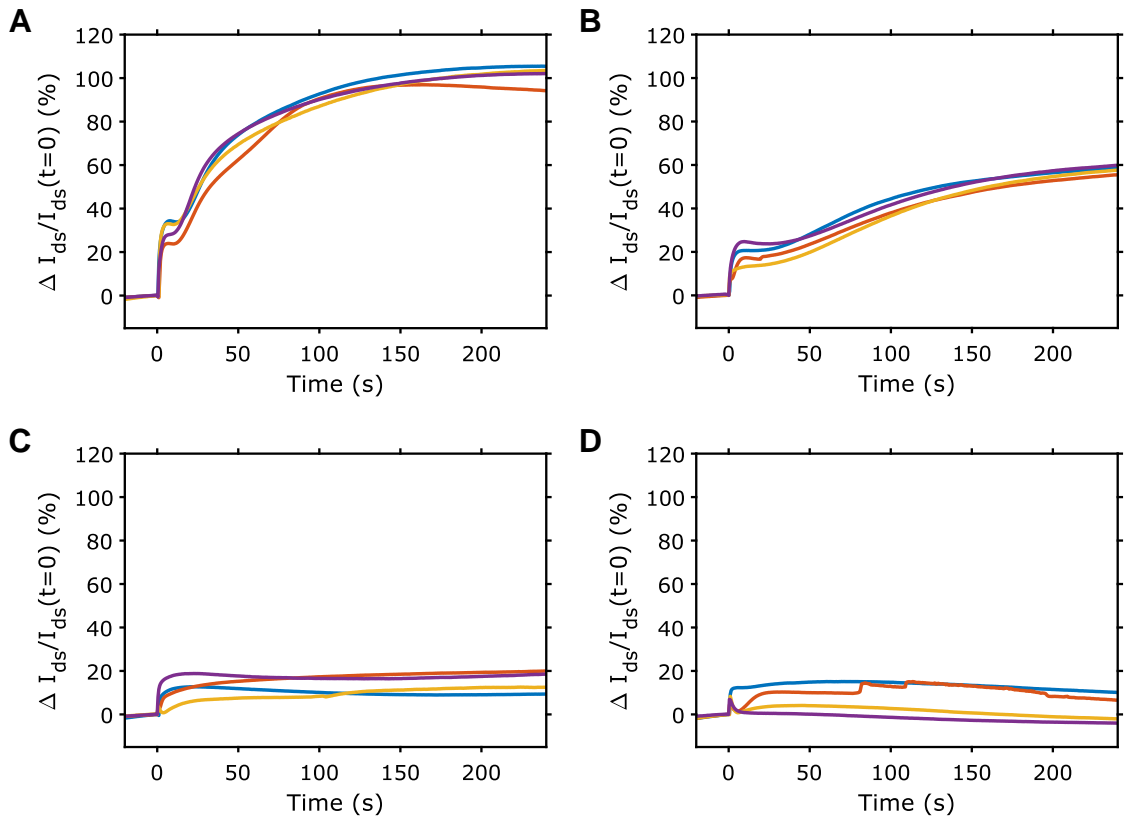

**Figure S1:** The transient change in $I_{ds}$ upon deposition of NaCl solutions, showing trials for individual devices in different colours. **(A)** 1 M, **(B)** 100 mM, **(C)** 10 mM, and **(D)** 0 mM (deionised water).

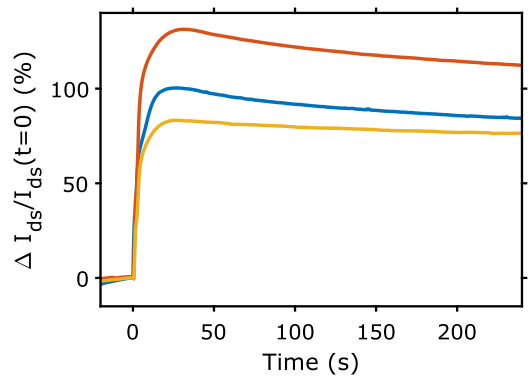

**Figure S2:** The transient change in $I_{ds}$ upon deposition of 1 M NaCl solution, onto a HIFET pre-hydrated with 8 µL of deionised water >2 hours prior to testing. Trials for individual devices shown in different colours.

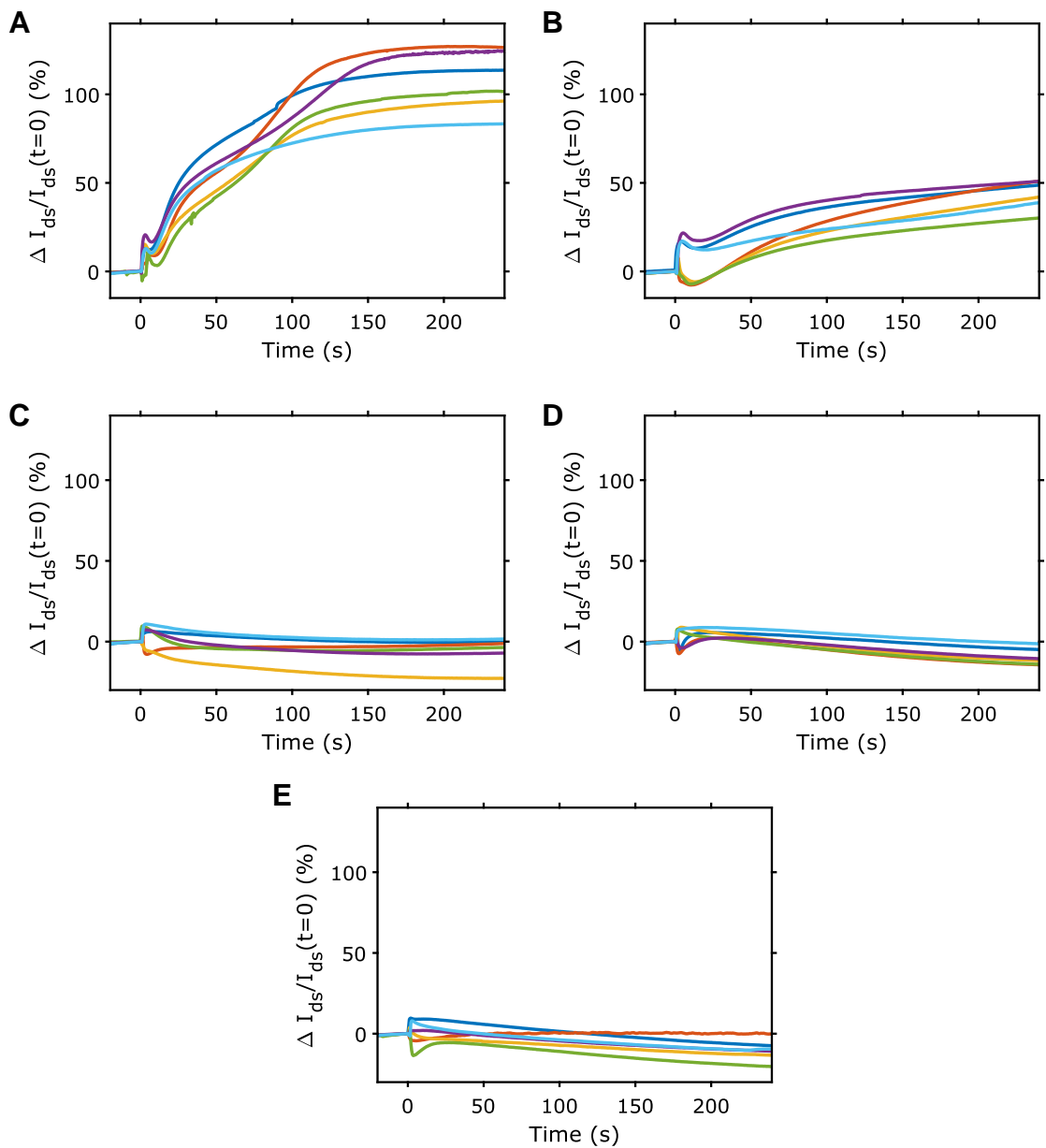

**Figure S3:** The transient change in $I_{ds}$ upon deposition of KCl solutions, showing trials for individual devices in different colours. **(A)** 1 M, **(B)** 100 mM, **(C)** 10 mM, **(D)** 1 mM, and **(E)** 0 mM (deionised water).

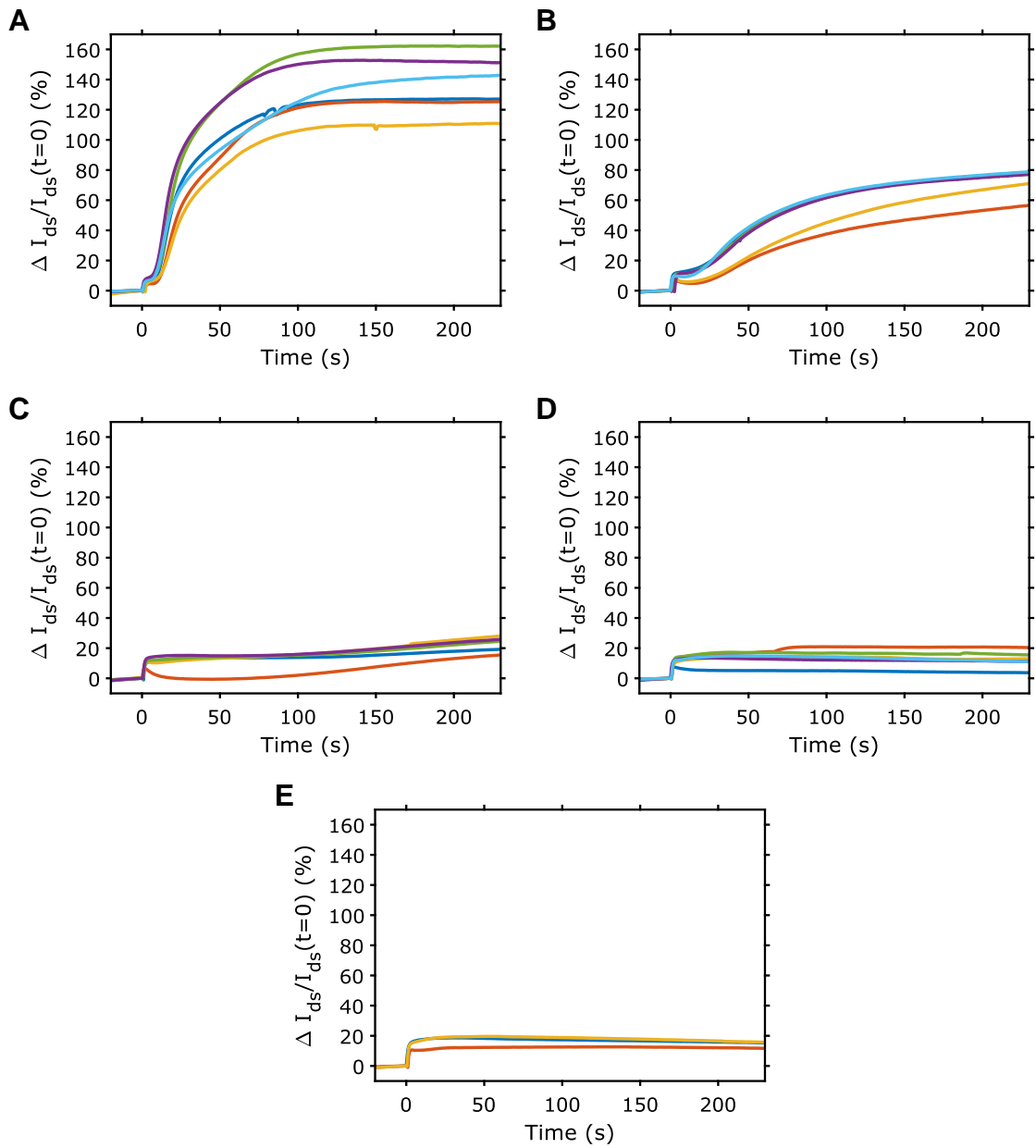

**Figure S4:** The transient change in $I_{ds}$ upon deposition of HCl solutions, showing trials for individual devices in different colours. **(A)** pH 0, **(B)** pH 1, **(C)** pH 2, **(D)** pH 3, and **(E)** pH 7 (deionised water).

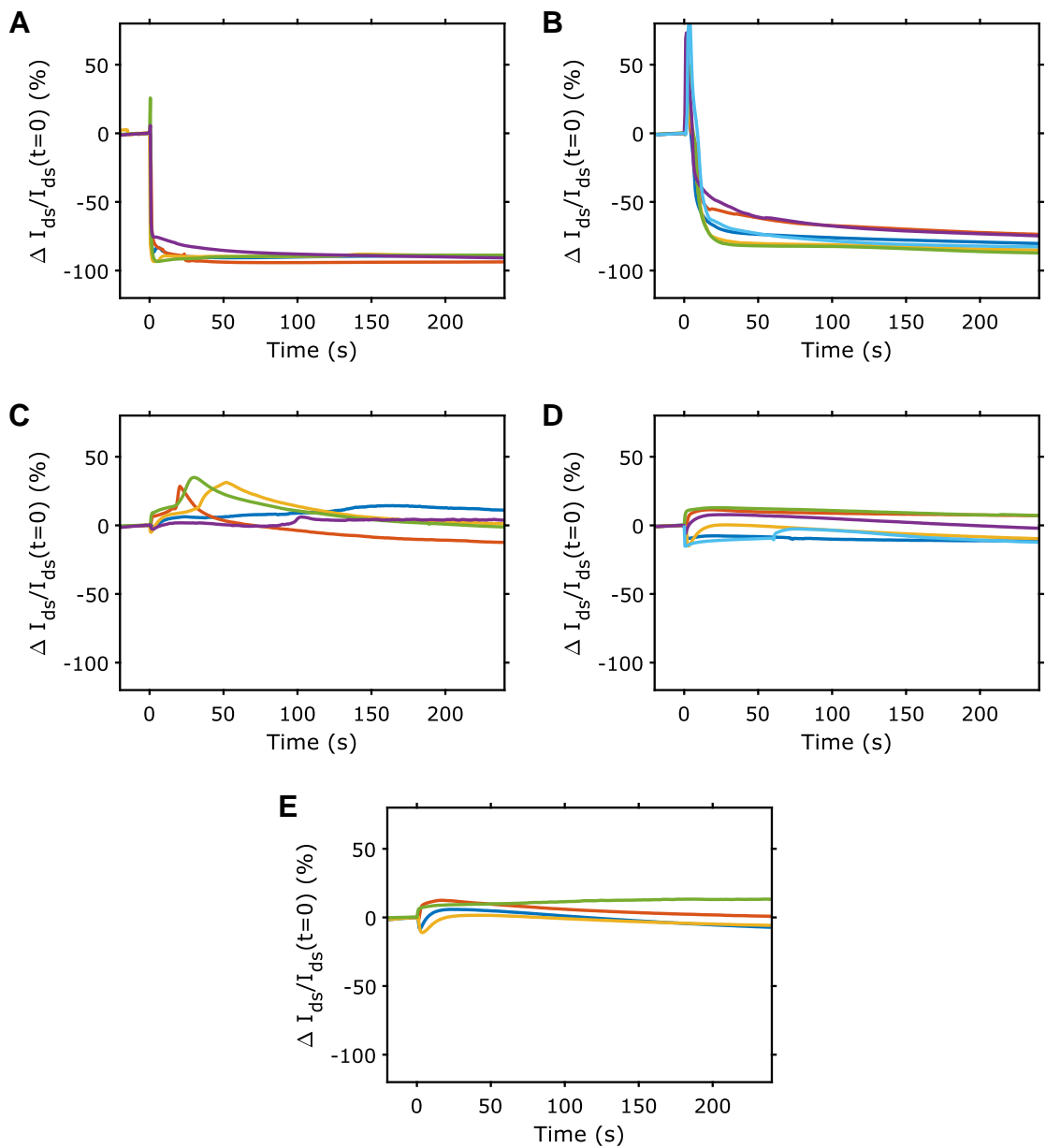

**Figure S5:** The transient change in $I_{ds}$ upon deposition of NaOH solutions, showing trials for individual devices in different colours. **(A)** pH 14, **(B)** pH 13, **(C)** pH 12, **(D)** pH 11, and **(E)** pH 7 (deionised water).

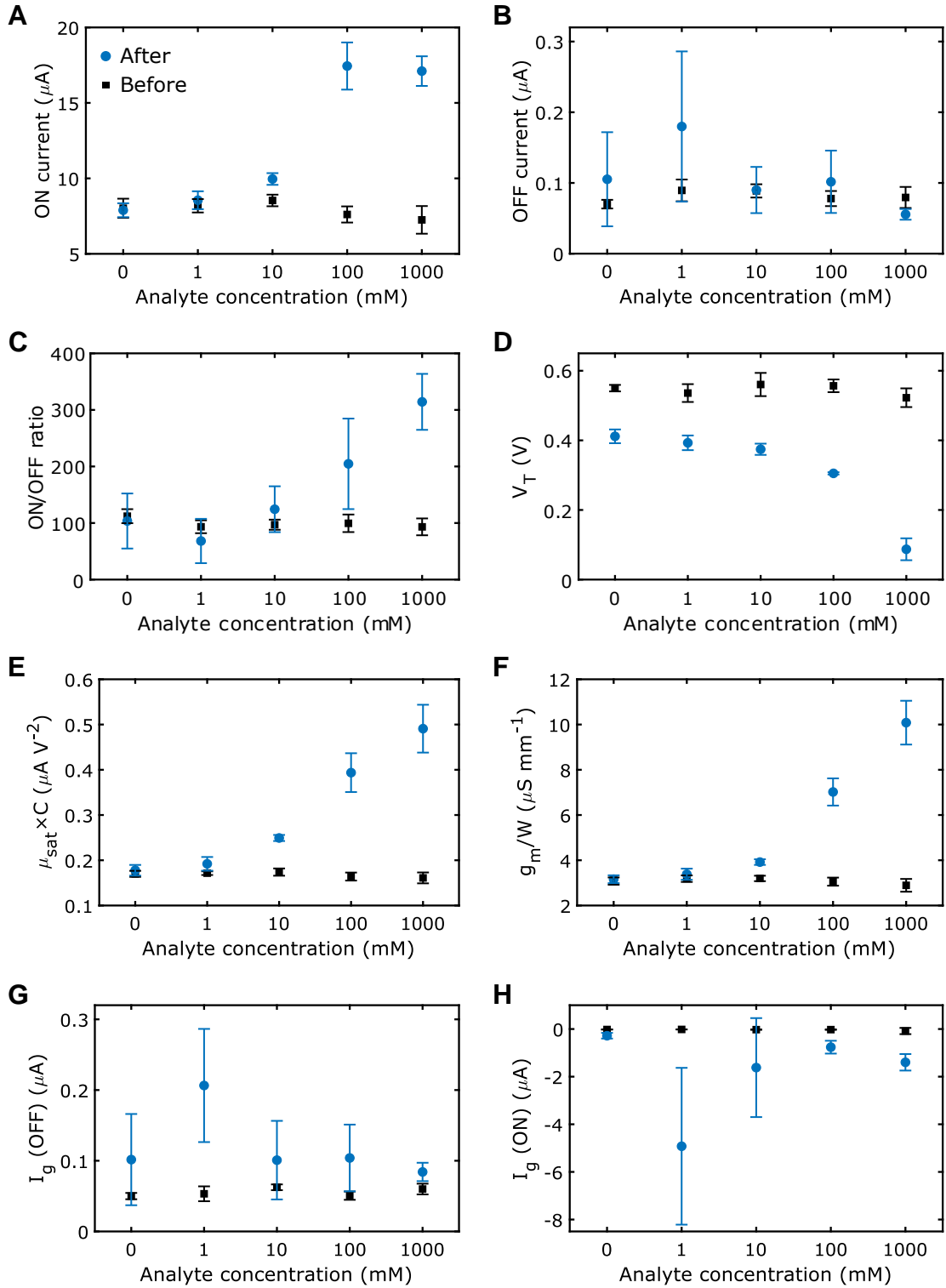

**Figure S6:** Average figures of merit for HIFETs in the saturation regime before and after deposition with KCl solutions of different concentrations. **(A)** ON current (maximum $I_{ds}$ in transfer sweep). **(B)** OFF current (minimum $I_{ds}$ in transfer sweep). **(C)** ON/OFF ratio. **(D)** Estimated threshold voltage ($V_T$). **(E)** Product of saturation mobility ($\mu_{sat}$) and capacitance (C). **(F)** Maximum transconductance ($g_m$). **(G)** Gate current ($I_g$) in OFF state (minimum $I_{ds}$). **(H)** Gate current ($I_g$) in ON state (maximum $I_{ds}$). An abbreviated version of this figure is given as Figure 3 in the main text.

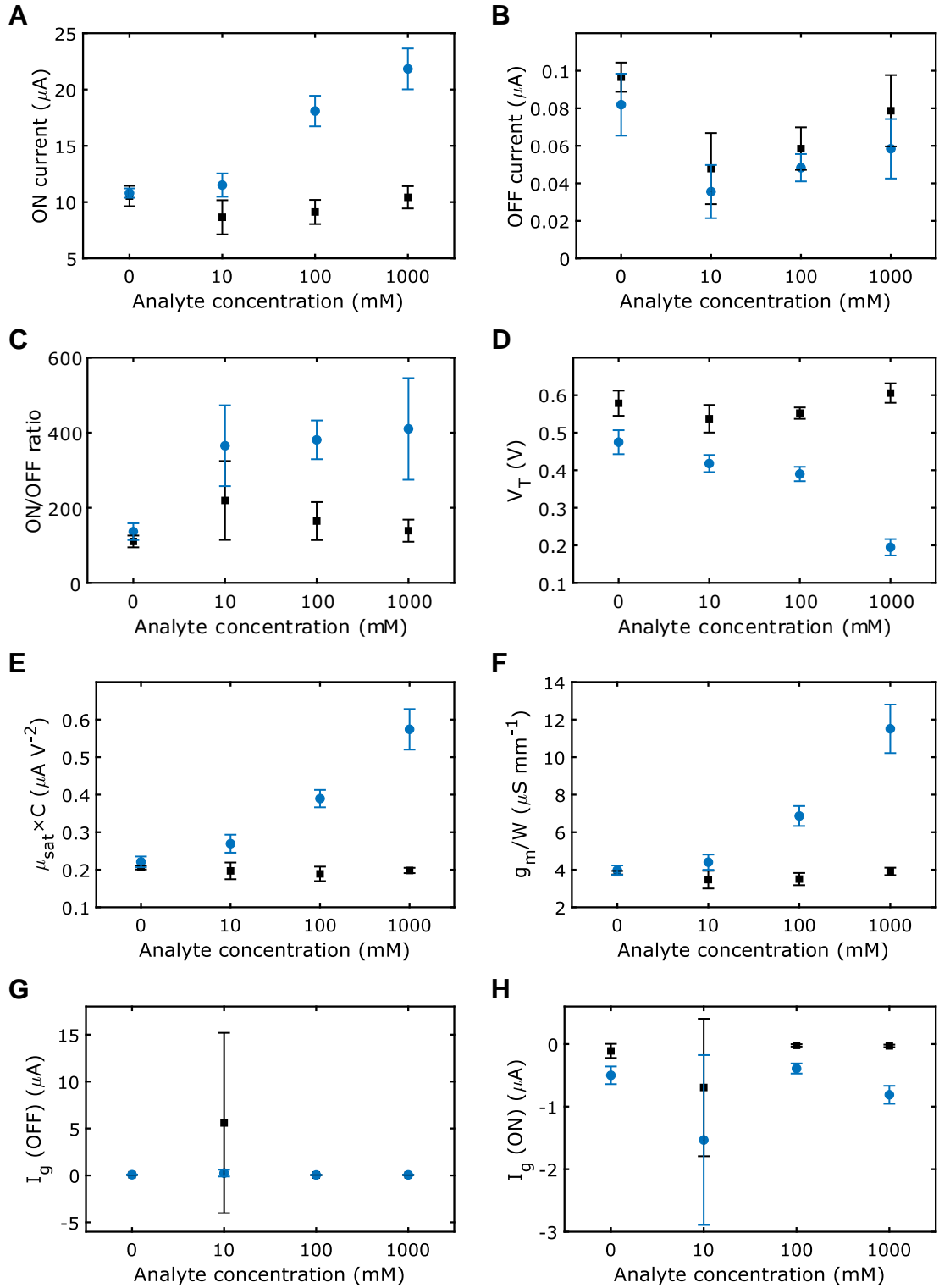

**Figure S7:** Average figures of merit for HIFETs in the saturation regime before and after deposition with NaCl solutions of different concentrations. **(A)** ON current (maximum $I_{ds}$ in transfer sweep). **(B)** OFF current (minimum $I_{ds}$ in transfer sweep). **(C)** ON/OFF ratio. **(D)** Estimated threshold voltage ($V_T$). **(E)** Product of saturation mobility ($\mu_{sat}$) and capacitance (C). **(F)** Maximum transconductance ($g_m$). **(G)** Gate current ($I_g$) in OFF state (minimum $I_{ds}$). **(H)** Gate current ($I_g$) in ON state (maximum $I_{ds}$).

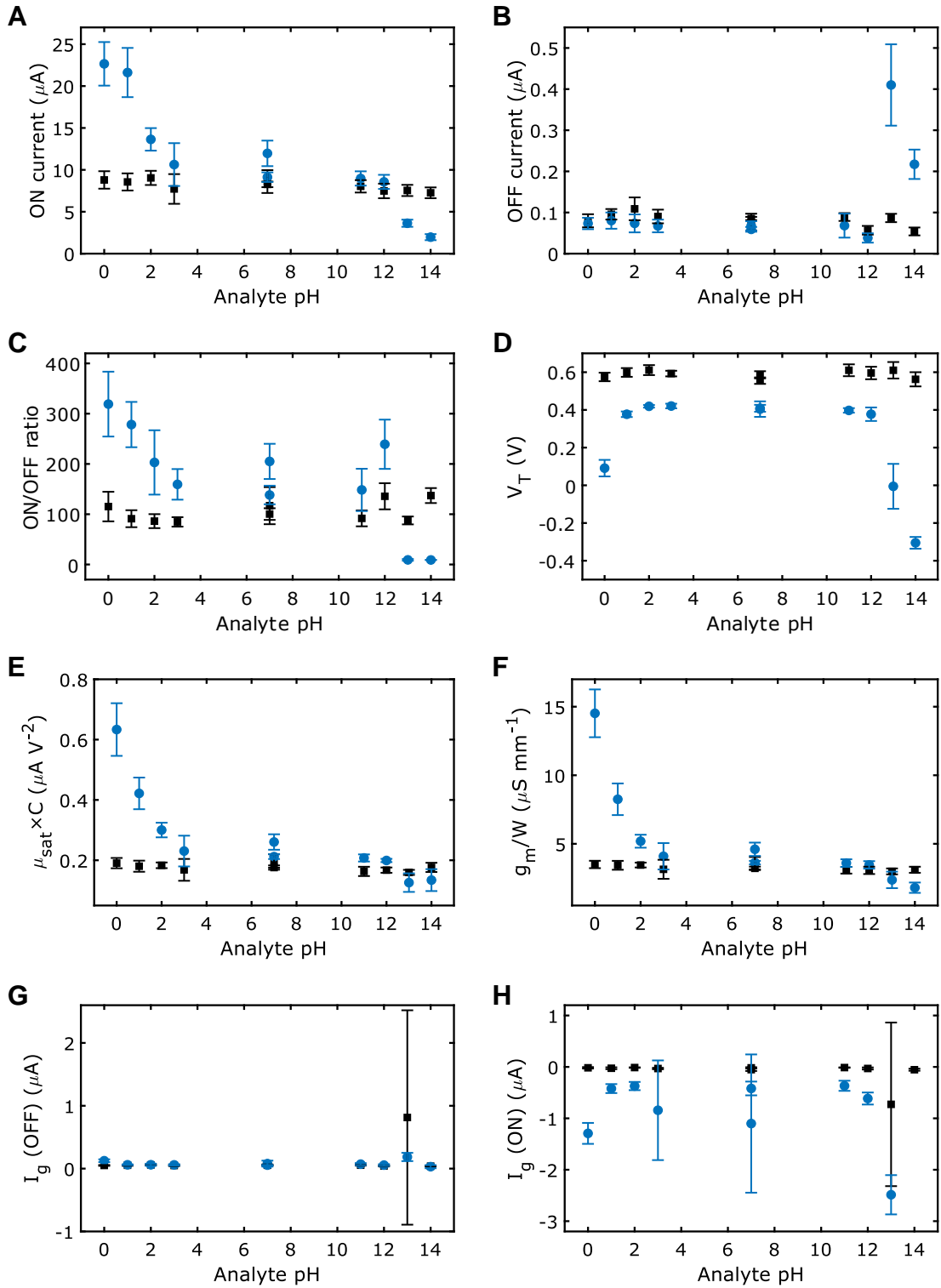

**Figure S8:** Average figures of merit for HIFETs in the saturation regime before and after deposition with HCl and NaOH solutions of different concentrations. **(A)** ON current (maximum $I_{ds}$ in transfer sweep). **(B)** OFF current (minimum $I_{ds}$ in transfer sweep). **(C)** ON/OFF ratio. **(D)** Estimated threshold voltage ($V_T$). **(E)** Product of saturation mobility ($\mu_{sat}$) and capacitance (C). **(F)** Maximum transconductance ($g_m$). **(G)** Gate current ($I_g$) in OFF state (minimum $I_{ds}$). **(H)** Gate current ($I_g$) in ON state (maximum $I_{ds}$).

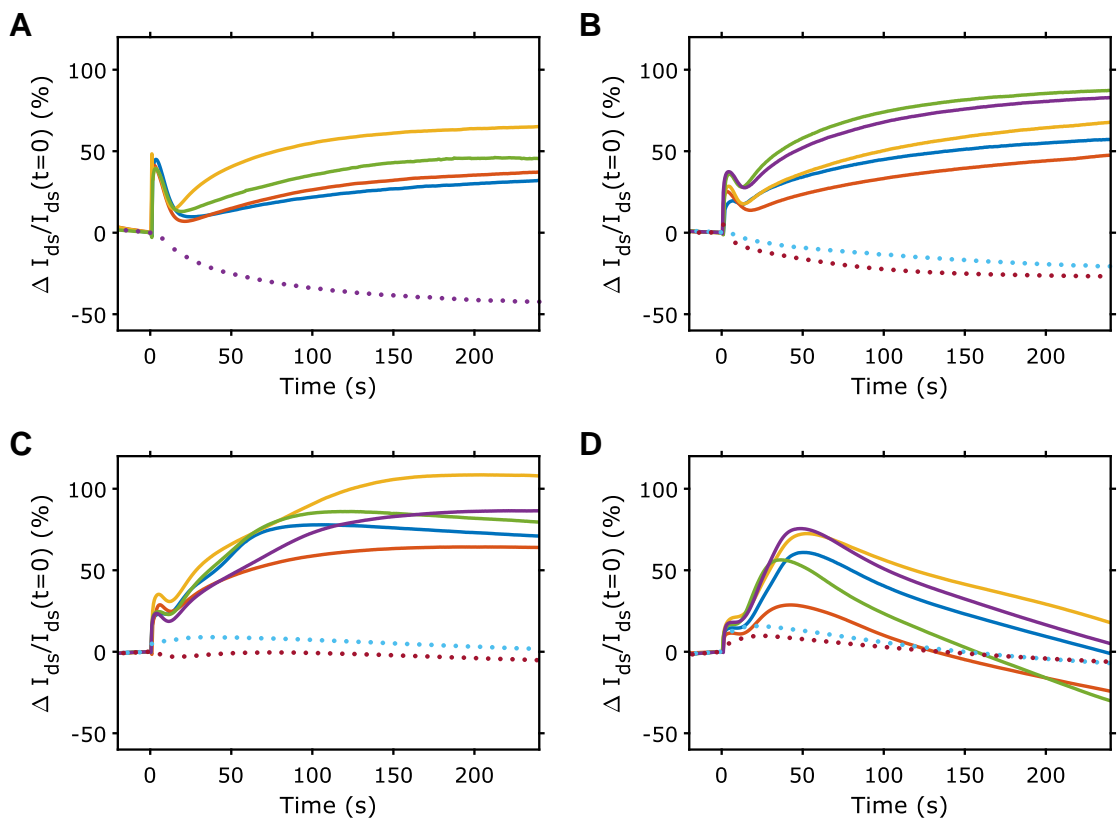

**Figure S9:** The transient change in $I_{ds}$ upon deposition of 1 M NaCl solutions at different $V_g$, showing trials for individual devices in different colours. **(A)** $V_g$ = +0.3 V, **(B)** $V_g$ = 0 V, **(C)** $V_g$ = -0.3 V, and **(D)** $V_g$ = -0.6 V.

8